# Annealing Effect on Microstructure and Superconductivity of $K_xFe_ySe_2$ Synthesized by an Easy One-step Method

W. Zhou, X. Li, X. Zhou, F. F. Yuan, Y. Ding, J. C. Zhuang, Y. Sun, H. L. Zhou, Z. X. Shi [a]

*Department of Physics and Key Laboratory of MEMS of the Ministry of Education, Southeast University, Nanjing 211189, China*

[a] E-mail Address: zxshi@seu.edu.cn

## Abstract

High-quality superconducting $K_xFe_ySe_2$ single crystals were synthesized using an easy one-step method. Detailed annealing studies were performed to make clear the phase formation process in $K_xFe_ySe_2$. Compatible observations were found in temperature-dependent X-ray diffraction patterns, back-scattered electron images and corresponding electromagnetic properties, which proved that good superconductivity performance was close related to the microstructure of superconducting component. Analysis based on the scaling behavior of flux pinning force indicated that the dominant pinning mechanism was $\Delta T_c$ pinning and independent of connectivity. The annealing dynamics studies were also performed, which manifested that the humps in temperature-dependent resistance ($RT$) curves were induced by competition between the metallic/superconducting and the semiconducting/insulating phases.

**Keywords:** annealing, flux pinning, phase separation.

## 1. Introduction

$K_xFe_ySe_2$ [1] was one of the recent discovered iron-based superconductors [2] (IBSs) which gave rise to extensive studies [3-5]. Crystallized as I4/mmm structure (122 type) [6], $K_xFe_ySe_2$ has the same tetragonal FeSe layer, from which superconducting carriers flow, as the Fe(Te, Se) compounds [7, 8] with the simplest crystal structure in IBSs. Fe(Te, Se) compounds bear nearly isotropic upper critical field ($H_{c2}$) and rather high critical current density ($J_c$) [9], but its low superconducting transition temperature ($T_c$) restricts the potential application. Intercalation of K can effectively enhance the $T_c$ of pure FeSe from 8 K [8] to about 32 K. However, due to the high activity of K element, complex microstructures with various components were formed, which leads to the small superconducting shielding volume and critical current density $J_c$ [10].

Previous studies pointed out that rich phenomena such as, Fe vacancy [4], double magnetic penetration [11] and phase separation [12-14], exist in $K_xFe_ySe_2$. Phase separation added the difficulty in solving many fundamental questions. Primarily, the parental phase and the actual superconducting component are still controversial. Many compounds of different compositions were proposed as the parental phase [15-18]. However, no widely perceived conclusion reached. Beside these important issues that wait to be addressed, the phase formation process should be made clear first. In fact, reversible tuning of superconducting phase in the present crystal via post annealing and quenching method [5, 19] has been reported by many groups, which leads to the subsequent observation of inhomogeneous superconducting network [16, 20]. Inhomogeneous superconductivity has been investigated exhaustively for various configurations of special structures in many systems that grow purposely [21, 22]. However, few studies of superconducting network that forms naturally were reported. It is important to make clear the formation process of this regular superconducting network and study the corresponding evolution of its physical properties. In this report, $K_xFe_ySe_2$ single crystal was synthesized by an easy one-step self-flux method. The annealing effects on microstructure and superconductivity as well as the annealing dynamics were studied systematically.

## 2. Sample Preparation and Post Annealing Method

Single crystals used in this report were synthesized by an easy one-step method. First, pure K, Fe and Se elements with mole ratio of 0.8: 2: 2 were put into alumina crucibles and sealed in vacuumed silica tubes with thickness thicker than 1.5 mm. A turbo pump was used to vacuum the silica tubes to pressure inside around $8.5 \times 10^{-5}$ mbar simultaneously in the process of sealing. And then the silica tubes were put into high temperature box-furnace and heated to 900 °C in 2 hours. The highest sintering temperature can't exceed 900 °C for the easy reaction between K and the silica tubes. After holding at 900 °C for about 3h, the furnace was set to cooling at rate of -3 °C/h to 780 °C, and then furnace cooled to room temperature. The short sintering time is remarkably superior to the previous synthesis methods [10, 23]. The as grown samples (hereafter noted as AG) are easily cleaved into shining plates along the crystalline *c*-axis direction. For the annealing process, the AG sample was first cleaved into slices and sealed in many small vacuumed tubes using the same method mentioned above. And then the small tubes were heated to different temperatures in an hour, held for various times and quenched in air. For every measurement that will be discussed subsequently, the annealed and quenched samples (hereafter noted as AQ) are selected from a same batch and heated in the same box-furnace.

## 3. Annealing Effect on Basic Sample Characterizations

Temperature-dependent powder X-ray diffraction (XRD), with Cu Kα radiation from 10° to 70°, was carried out for the AG crystal. The XRD patterns were shown in figure 1. Almost all the main peaks can be indexed with I4/mmm space group, consistent with ref. [10, 24]. During the heating process, XRD measurements were performed 15 minutes later when the temperature reached at the setting points. As can be seen, the sample was badly crystallized in the beginning. However, great change possibly originated from sample re-crystallization occurred at temperature about 500 °C. For the cooling process, measurements was done once the temperature reached at the setting points with cooling rate of about 20 ~ 30 °C/min. Figure 1 (b) shows the enlarged view of the (103) peak whose intensity is relatively higher. Two peaks were clearly observed when heating the sample below 500 °C. Previous studies suggest that the former peak labeled by club represents the superconducting phase [16]. As can be seen, for AQ of 300/400 °C, higher peaks for the superconducting phase were witnessed. Therefore, temperature around 300/400 °C may be a suitable growth condition for the superconducting phase.

In figure 2, back-scattered electron images of SEM measurements were shown. As can be seen, for the AG and 200 °C AQ samples, large and disconnected rectangular bars were observed, which become finer and form well-connected network at annealing temperature of 300 °C or higher. The densest network was obtained for samples annealed and quenched at 300 or 400 °C. For 400 and 500 °C annealed samples, large amounts of cracks can be seen, which is very bad for obtaining good superconductivity performance. Such phase separation phenomena were also reported in ref. [16, 20], and the rectangular bars or networks were suggested to be responsible for superconductivity. Simply considering the distribution of superconducting component, good superconductivity performance may only be obtained for AQ of 300/400 °C.

## 4. Annealing Effect on Magnetic Properties

Annealing effect on properties that directly associated with superconductivity was first revealed through magnetic measurement, which was performed via the VSM option of a Physical Property Measurement System (PPMS). Figure 3 shows the temperature dependent magnetization (*MT*) curves for samples annealed at different temperatures. As can be seen, the superconducting transition temperature $T_c^{Mag}$ is around 31 K and unchanged for different annealing temperatures. As pointed out in ref. [25], the superconducting fraction can't be obtained directly from *MT* measurement due to the serious influence of connectivity for inhomogeneous superconductors. Therefore, to compare the annealing effect on superconductivity performance for these AQ samples, field-dependent magnetization (*MH*) should be measured. In figure 4, *MH* loops and corresponding $J_c$ calculated by Bean model [26] at 5 K for both the AG and AQ samples were shown. *MH* loops for the AG sample are small and asymmetric (*MH*s for other measured temperatures were not shown), which leads to a peak in low field region in $J_c$ (*B*) curves. Symmetric *MH* loops are observed even annealing and quenching sample at 200 °C, which means bulk pining dominated even though the rectangular bars are not well connected as shown in the SEM back-scattered electron images. The highest $J_c$ is obtained for sample annealed at 300 °C. At higher measured temperature, $J_c$ for 300 °C AQ sample dropped below that of 400/500 °C AQ samples in high fields region, as shown in figure 5. Generally, $J_c$ is composed by intra-granular and inter-granular parts for polycrystal [27]. Weak links exist for inter-granular. In case of existence of large numbers of weak links, for higher measured temperature, $J_c$ from the inter-granular part should be much more sensitive to magnetic field, which thus causes quick decrease of whole $J_c$. The situation may be also suitable to the case of $K_xFe_ySe_2$ single crystals with nonuniform superconducting component.

To further investigate whether the flux pinning mechanism is affected by the structure of superconducting component, the scaling behavior of $f_p(h)$ is analyzed. Here $f_p = F_p / F_p^{max}$,

$h = H/H_P^{max}$, and $F_p = \mu_0 H J_c$ is the vortex pinning force. According the previous studies [28], the scaling behavior of $f_p(h)$ can often be analyzed by the following three formulas:

$$f(h) = 3h^2(1-\frac{2h}{3}), \qquad \text{for } \Delta\kappa \text{ pinning} \qquad (1)$$

$$f(h) = \frac{9}{4}h(1-\frac{h}{3})^2, \qquad \text{for } \Delta T_c \text{ pinning} \qquad (2)$$

$$f(h) = \frac{25}{16}\sqrt{h}(1-\frac{h}{5})^2, \qquad \text{for surface pinning} \qquad (3).$$

As can be seen in figure 6, $f_p(h)$ for different measured temperatures can be scaled together, indicating the same pinning mechanism. Typical $\Delta T_c$ pinning behavior, consistent with ref. [5], was observed, which means the core normal point-like pinning mechanism is dominant. For samples annealed by different temperatures, the pinning types are the same, which is in sharp contrast with the quite different configurations of superconducting structures as shown in the inset of figure 6.

## 5. Annealing Effect on Transport Properties

Temperature dependences of resistance for the annealed single crystals were measured on a home-made low temperature measurement equipment (Keithley 2400: current source, Keithley 2182: nanovoltmeter, LakeShore 331: temperature controller) using the standard four-probe method for the easy preparation of electrodes. Figure 7 shows the temperature-dependent resistance for samples annealed at different temperatures. The AQ sample at 300 °C for an hour exhibits the best superconducting transition which is consistent with XRD, SEM and *MH* results. The AQ sample at 400/500 °C doesn't show good superconductivity performance as magnetic measurement. This phenomenon may be caused by the cracks observed in SEM images for high temperature AQ samples. For magnetic measurement, since the superconducting loops exist still, large superconducting signals can be measured easily. These results indicate 300 °C annealing and quenching may be a more safe way to obtain superconductivity of good performance.

Therefore, we annealed the AG samples at 300 °C for various times to study the dynamic process of annealing. As can be seen in figure 8, great improvement of superconductivity was obtained through annealing at 300 °C for only 30 minutes. No zero resistance was observed for the sample annealed at 300 °C for 5 hours, and the low temperature resistance even exhibited semiconducting behavior. Long time annealing seemed to be harmful to superconductivity. In addition, it's noticeable that there exist of humps in all *RT* curves, which has been demonstrated irrelevantly to magnetic or structure phase transition [29]. In figure 9 one can see, the $T_c$, including both $T_c^{onset}$ and $T_c^{zero}$, does not change much for different annealing times, which may indicate no change of superconducting component though the connectivity may be clearly different. The temperature position of hump in *RT* curves and the slope just above $T_c$ were plotted in the same figure. The slopes were obtained by linear fitting of normalized resistance ($R/R_{35\,K}$) data from $T_c^{onset}$ to 35 K. As can be seen, both the hump position and slope value show similar trends with the annealing time, first increasing and then decreasing. The situation may happen because of coexistence of semiconducting/insulating and metallic phases in the present samples. The value of slopes reflects the proportion between metallic and semiconducting/insulating components. When the proportion of metallic phase is larger, the hump position moves to higher temperature and the slope value becomes smaller. Since the best superconductivity performance is obtained by annealing sample for 1h with the highest hump temperature and smallest slope value, one can naturally suppose that the normal state of superconducting phase is metallic. The same trend was also reflected in *RT* curves of samples annealed at different temperatures. Therefore, the hump may be resulted from competition between the metallic and semiconducting phases. When the annealing time is less than 1 h, the proportion of the metallic phase (superconducting at low temperatures) increases with annealing time, and leads to higher hump temperature as well larger slope value. However, another semiconducting/insulating component, which may be caused by either cracks or re-crystallization, grows quickly for more than 1 hour's annealing. The growth of semiconducting/insulating phase leads to the hump position and slope moving to low temperatures and low values again.

## 6. Conclusion

In summary, detailed annealing study has been performed on single crystals synthesized by an easy one-step method. Back-scattered electron images revealed that well-connected superconducting network can be formed by annealing and quenching samples at temperatures above 300 °C. Accompanied with the formation of superconducting networks, superconductivity of high $J_c$ can be obtained easily. However, both high temperature and long time annealing were demonstrated to be harmful to the growth of superconducting phase. Detailed annealing-temperature-dependent and time-dependent transport results suggested that the superconducting phase should be metallic in the normal state and the competition between metallic and semiconducting/insulating phases aroused the humps in *RT* curves.

**Acknowledgments**: This work was supported by the Natural Science Foundation of China, the Ministry of Science and Technology of China (973 project: No. 2011CBA00105), and Jiangsu Science and Technology Support Project (Grant No. BE 2011027).

# Figure Captions

Figure 1 (a) Temperature-dependent powder XRD patterns. (b) The partial enlarged view of (a).

Figure 2 Back-scattered electron images of the AG and AQ samples with two magnified scales (20000x and 5000x). Disconnected rectangular bars for low temperature annealing (AG, 200 °C) and well-connected networks for high temperature annealing (300 °C, 400 °C, 500 °C) were observed.

Figure 3 Temperature dependences of zero-field-cooled (ZFC) and field-cooled (FC) magnetization for AG and AQ samples with magnetic filed $H$ = 20 Oe ($H//c$).

Figure 4 (a) Magnetic hysteresis loops (MHLs) for samples annealed and quenched at different temperatures with the measured temperature of 5 K ($H//c$). (b) Critical current density $J_c$ calculated from MHLs for the AG and AQ samples.

Figure 5 $J_c$ (20 K) calculated from MHLs for AG and AQ samples of different annealing temperatures.

Figure 6 Temperature-dependent scaling behaviors of flux pinning force ($f_p(h)$) for 200 and 400 °C AQ samples. Inset: the back-scattered electron images.

Figure 7 Temperature dependence of resistance ($RT$) for AG and AQ samples of different annealing temperatures.

Figure 8 $RT$ curves for 300 °C AQ samples with different annealing times.

Figure 9 Hump temperatures, slopes, $T_c^{onset}$s, and $T_c^{zero}$s for 300 °C AQ samples vs the annealing time. The slopes are obtained by linear fitting of data above $T_c$ in $RT$ curves in the inset of figure 8.

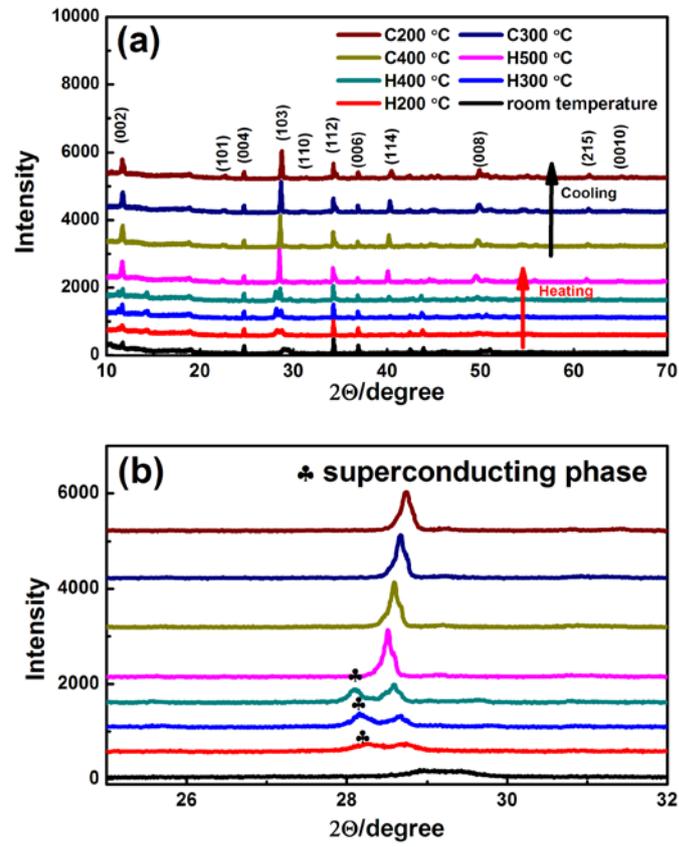

Figure 1 W. Zhou *et al.*

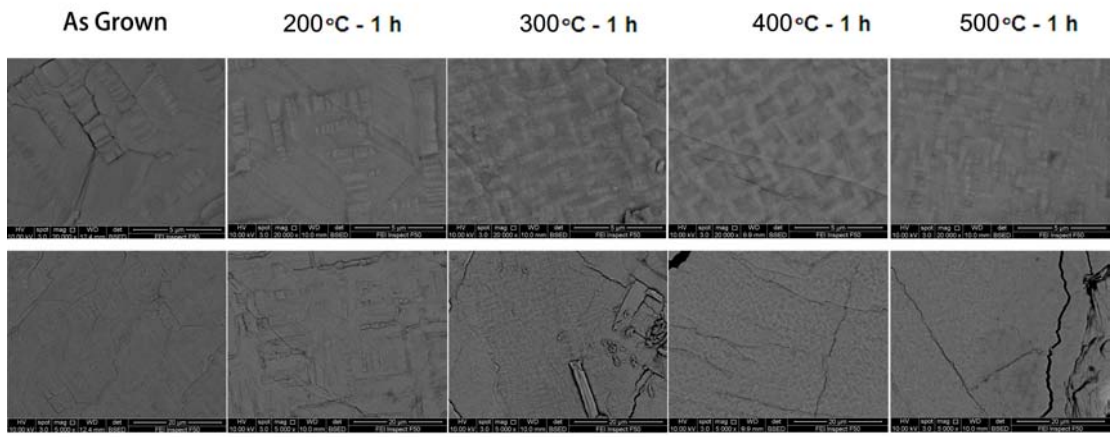

Figure 2 W. Zhou *et al*.

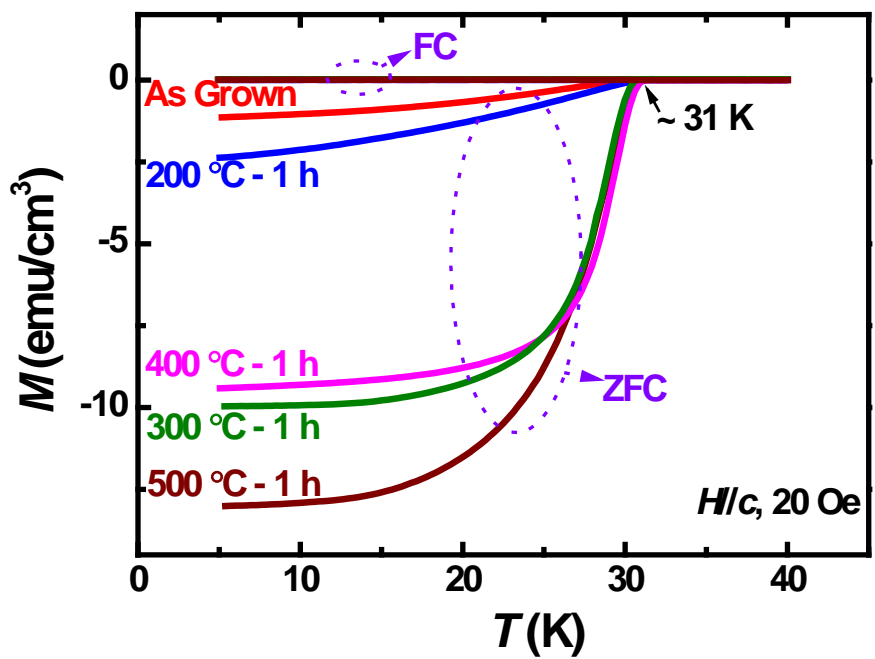

Figure 3 W. Zhou *et al*.

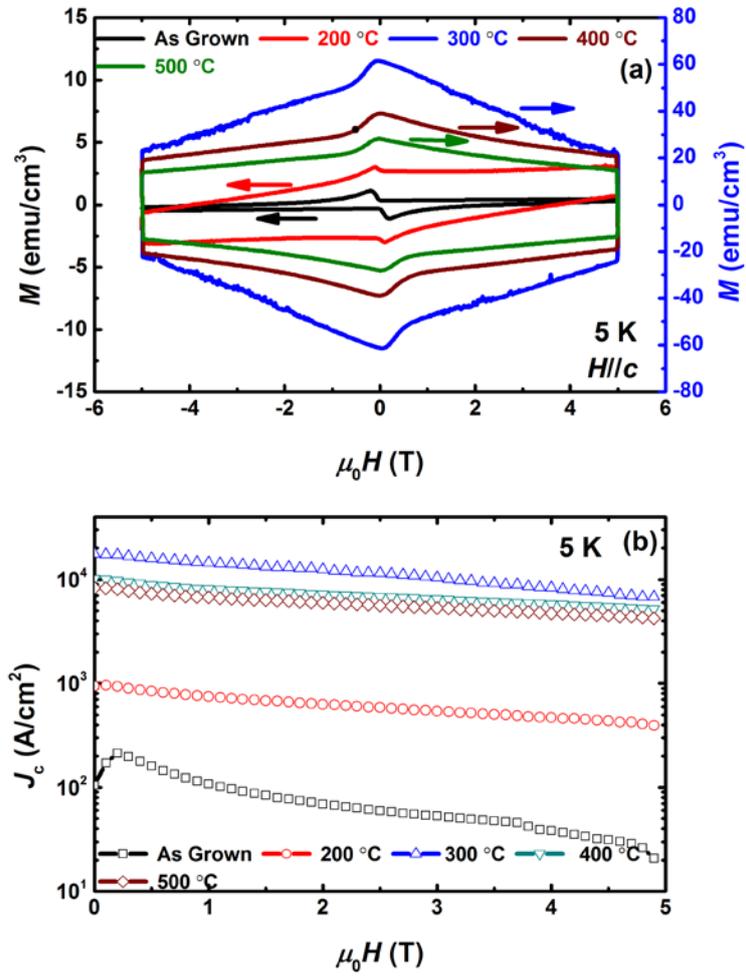

Figure 4 W. Zhou *et al.*

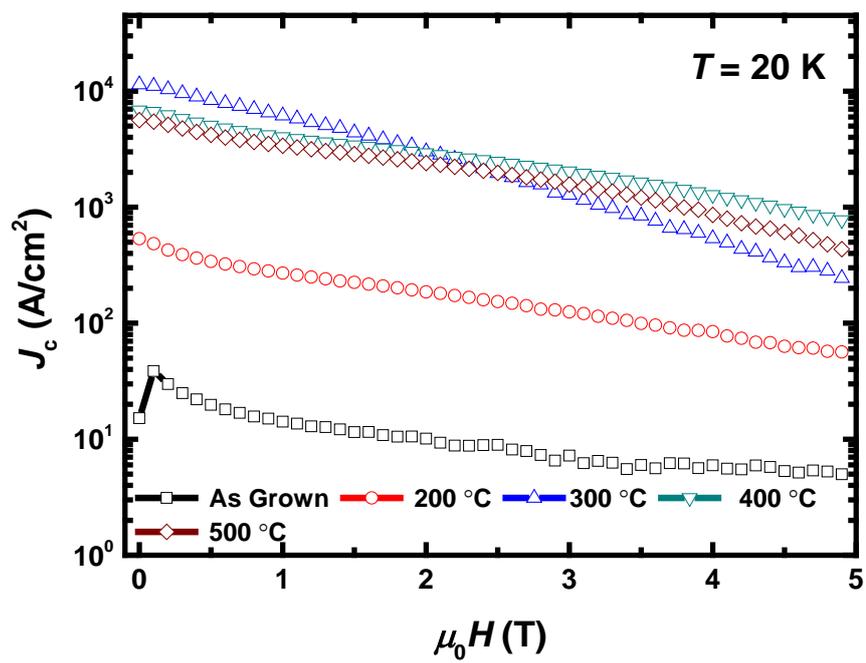

Figure 5 W. Zhou *et al.*

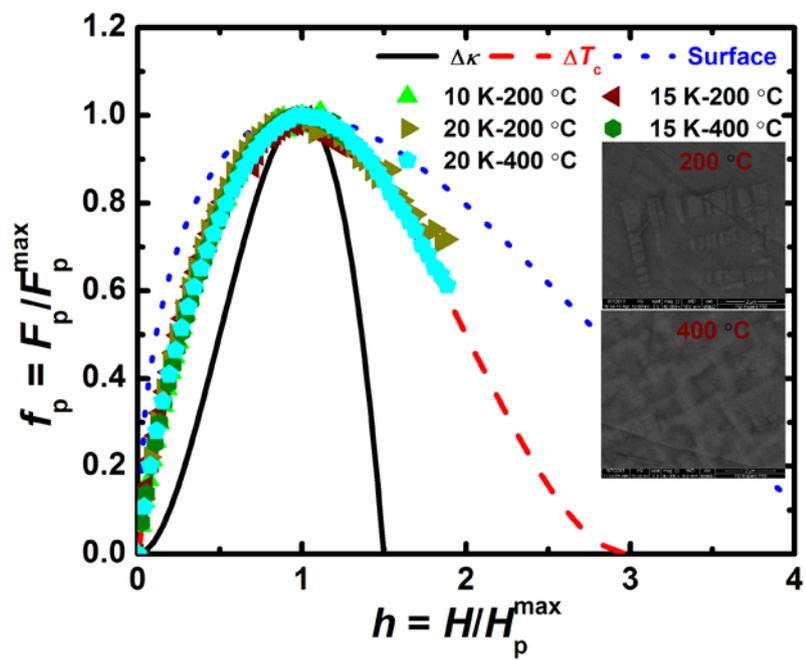

Figure 6 W. Zhou *et al.*

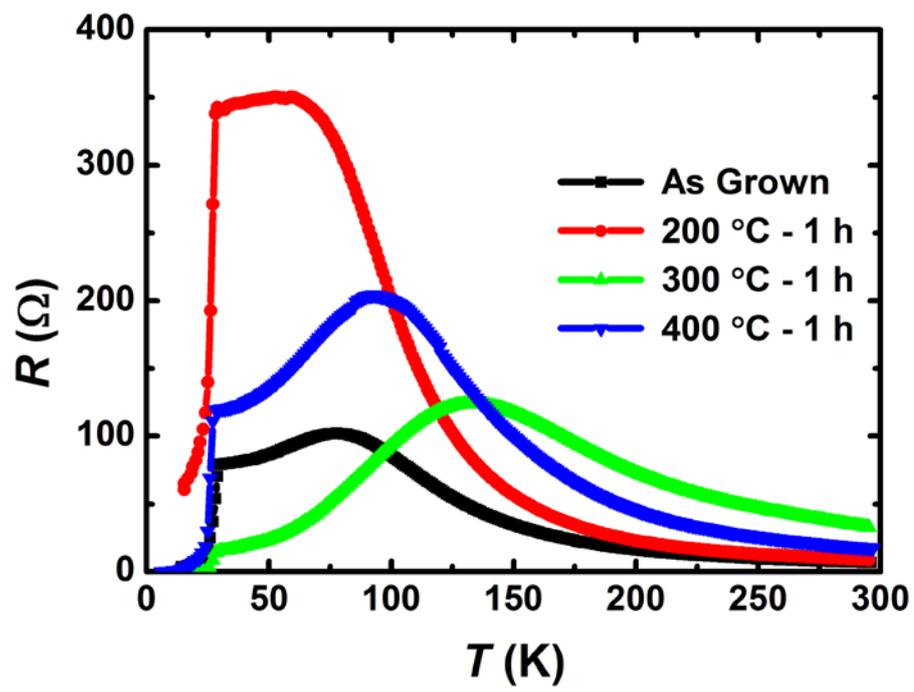

Figure 7 W. Zhou *et al.*

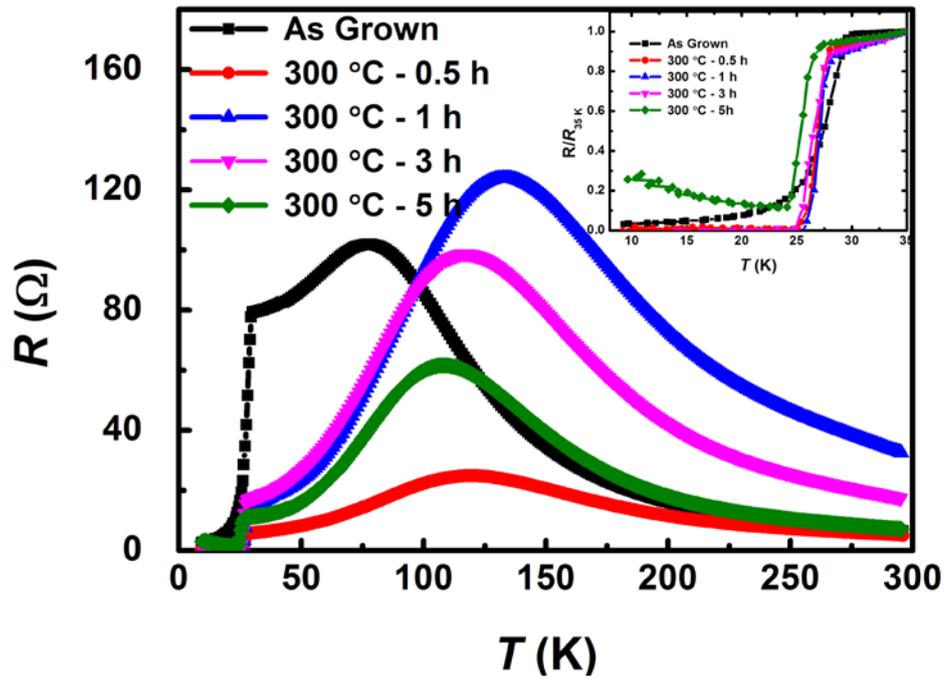

Figure 8 W. Zhou *et al.*

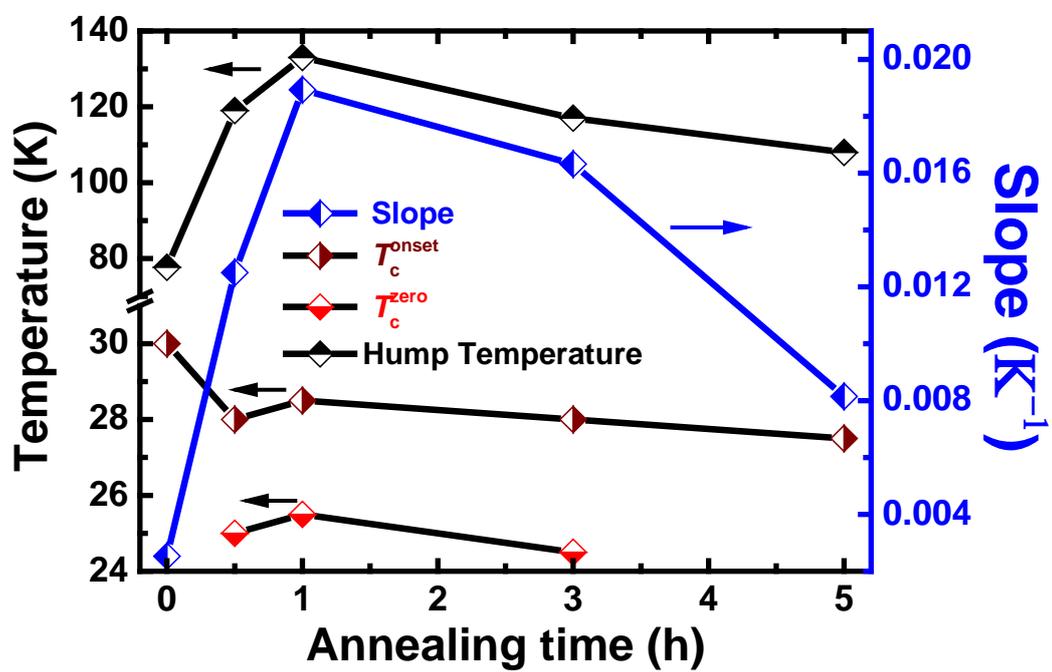

Figure 9 W. Zhou *et al.*